\documentclass[preprint2]{aastex}
\bibpunct{(}{)}{;}{a}{}{,}
\begin{document}


\shorttitle{Chemical Signature of the First Stars}
\shortauthors{Karlsson et al.}

\title{Uncovering the Chemical Signature of the First Stars in the Universe}
\author{Torgny Karlsson\altaffilmark{1}, Jarrett L. Johnson\altaffilmark{2}, and Volker Bromm\altaffilmark{2}}

\email{karlsson@nordita.org; jljohnson@astro.as.utexas.edu; vbromm@astro.as.utexas.edu}

\altaffiltext{1}{NORDITA, AlbaNova University Center, SE-106~91 Stockholm, Sweden}
\altaffiltext{2}{Department of Astronomy, University of Texas, Austin, TX 78712, USA}

\begin{abstract}
The chemical abundance patterns observed in metal-poor Galactic halo stars contain the signature of the first supernovae, and thus allows us to probe the first stars that formed in the universe.  We construct a theoretical model for the early chemical enrichment history of the Milky Way, aiming in particular at the contribution from pair-instability supernovae (PISNe).  These are a natural consequence of current theoretical models for primordial star formation at the highest masses.  However, no metal-poor star displaying the distinct PISN signature has yet been observed.  We here argue that this apparent absence of any PISN signature is due to an observational selection effect.  Whereas most surveys traditionally focus on the most metal-poor stars, we predict that early PISN enrichment tends to `overshoot', reaching enrichment levels of $[\mathrm{Ca}/\mathrm{H}]\simeq -2.5$ that would be missed by current searches.  We utilize existing observational data to place constraints on the primordial initial mass function (IMF).  The number fraction of PISNe in the primordial stellar population is estimated to be $<0.07$, or $\lesssim 40\%$ by mass, assuming that metal-free stars have masses in excess of $10~\mathcal{M_{\odot}}$.  We further predict, based on theoretical estimates for the relative number of PISNe, that the expected fraction of second generation stars below $[\mathrm{Ca}/\mathrm{H}]=-2$ with a dominant (i.e., $>90\%$) contribution from PISNe is merely $\sim 10^{-4} - 5\times 10^{-4}$. The corresponding fraction of stars formed from gas exclusively enriched by PISNe is a factor of $\sim 4$ smaller.  With the advent of next generation telescopes and new, deeper surveys, we should be able to test these predictions.  
\end{abstract}

\section{Introduction} \label{intro}
Little is known about the first stars in the universe, the so-called Population~III (Pop~III henceforth).  Over which range of masses were they formed?  And what was their relative distribution by mass, i.e., their initial mass function (IMF)?  The apparent lack of metal-free stars in the Galactic halo suggests that star formation in metal-free gas should be skewed towards relatively high masses, as compared to the mass scale of $\lesssim 1~\mathcal{M_{\odot}}$ found in the solar neighborhood (e.g., Bond 1981\nocite{bond81}; Oey 2003\nocite{oey03}; Karlsson 2005\nocite{karlsson05}; Tumlinson 2006\nocite{tumlinson06}).  This finding is corroborated both by theoretical arguments (e.g., Larson \& Starrfield 1971\nocite{larson71}) and by numerical simulations (Bromm et al. 1999\nocite{bromm99}; Bromm et al. 2002\nocite{bromm02}; Abel et al. 2002\nocite{abel02}; Yoshida et al. 2006\nocite{yoshida06}) following the collapse of primordial baryonic matter inside the very first non-linear structures, virialized dark matter halos of $\sim 10^6~\mathcal{M_{\odot}}$, so-called minihalos. As a consequence of the inefficiency of primordial gas to cool to temperatures below $\sim$ $10^2$~K, these simulations indicate that the first stars were predominantly very massive objects with masses $\gtrsim 100~\mathcal{M_{\odot}}$.

It can be argued that star formation in minihalos, apart from being regulated by H$_2$ cooling, occurred under simplified conditions where the absence of magnetic fields, ultraviolet radiation, and a pre-established turbulent velocity field resulted in a unique metal-free stellar population (e.g., Bromm \& Larson 2004\nocite{bl04}; Glover 2005\nocite{glover05}).  For instance, in shock-compressed or photo-ionized primordial gas, the formation of the coolant HD is boosted, leading to a decrease in the typical mass of metal-free stars down to $\sim10~\mathcal{M_{\odot}}$ (Johnson \& Bromm 2006\nocite{jb06}).  Similarly, lower mass stars may also be formed in turbulent molecular clouds with a negligible magnetic field (Padoan et al. 2007\nocite{padoan07}).  The presence of a magnetic field could also, in the case of magnetized accretion disks, lead to smaller stellar masses as the accretion onto the proto-stellar core may be reduced by magnetically driven outflows (Silk \& Langer 2006\nocite{sl06}).

In spite of their peculiarity and apparent rarity, the first, very massive stars ultimately set the stage for subsequent star and galaxy formation.  If mass-loss through rotationally induced stellar winds was limited (Ekstr\"{o}m et al. 2006\nocite{ekstrom06}), a significant fraction of these stars was likely to explode as pair-instability supernovae (PISNe), leaving no remnant behind (Heger \& Woosley 2002\nocite{heger02}).  Interestingly, $\eta$ Carinae and the Pistol Star, both in the mass range $\sim100-200~\mathcal{M_{\odot}}$, are two examples of such very massive stars in our own Galaxy.  Moreover, the recent detection of a particularly energetic supernova (SN) explosion, SN~2006gy (progenitor mass estimated to $m \sim 150 \mathcal{M_{\odot}}$) in NGC 1260, indicates that stars may explode as PISNe, even at low redshift (Smith et al. 2007\nocite{smith07}).

Since very massive stars are extremely shortlived ($\lesssim 3$ Myr), primordial PISNe ($140\le m/\mathcal{M_{\odot}}\le 260$, Heger \& Woosley 2002\nocite{heger02}) would supposedly be the first stellar objects to enrich the interstellar medium (ISM) with metals.  As predicted by homogeneous chemical evolution models (e.g., Ballero et al. 2006\nocite{ballero06}), such an initial enrichment would establish a metallicity floor and imprint the chemical signature of PISNe in the early ISM, from which the first low-mass ($m$ $\la$ 1 M$_{\odot}$) stars later formed.  The unique chemical signature of the primordial PISNe should thus predominantly be retained in the oldest and, in particular, the most metal-poor stars belonging to the Galactic halo population.  To date, no such signature has been found (e.g., Christlieb et al. 2002\nocite{cetal02}; Cayrel et al. 2004\nocite{cayrel04}; Cohen et al. 2004\nocite{cohen04}; Barklem et al. 2005\nocite{barklem05}; Frebel et al. 2005\nocite{frebel05}).  This has been regarded as an indication that stars in the early universe with masses in excess of $100~\mathcal{M_{\odot}}$ were extremely rare (e.g., Tumlinson et al. 2004\nocite{tvs04}; Ballero et al. 2006\nocite{ballero06}), if not altogether absent.  However, we shall argue that true second generation stars which were formed from material predominantly enriched by primordial PISNe have metallicities significantly above those of the stars in the metal-poor tail of the Galactic halo metallicity distribution function (MDF), at variance with the classical picture described above. Furthermore, although the mass fraction of Pop~III stars ending up as PISNe may be significant, the fraction of low-mass stars bearing their chemical signature is small, which explains the absence of such stars in previous surveys.

The organization of the paper is as follows. In \S \ref{scenario} we consider the chemical enrichment by the first stars in its cosmological context, while the detailed chemical enrichment model and parameters are discussed in \S \ref{ece}. The results are presented in \S \ref{results}, and we conclude with a discussion of the implications in \S \ref{discussion}.

\section{The Cosmological Context}\label{scenario}
The end of the epoch known as the cosmic dark ages, when our universe witnessed the first sources of stellar light, is one of the final frontiers in modern cosmology. One indirect way to search for the first, Pop~III stars is to detect their contribution to the cosmic infrared background. Although tantalizing progress has been made (e.g., Kashlinsky et al. 2007\nocite{kashlinsky07}), we have to await the advent of next generation telescopes, such as the {\it James Webb Space Telescope (JWST)} and the {\it Atacama Large Millimeter/submillimeter Array (ALMA)} (e.g., Walter \& Carilli 2007\nocite{walter07}), to be able to resolve individual sources and their birth sites, and to place stronger constraints on the IMF and the star formation rate (SFR) in primordial galaxies.

A complementary approach is to search for Pop~III signatures in the local fossil record, i.e., to look for specific chemical abundance ratios in the atmospheres of old, low-mass stars in the Milky Way which may have formed from gas enriched with the ejecta from the first SNe.  This general strategy is termed near-field cosmology, and this particular approach is a powerful way to learn about the first generations of stars.  Unlike stars exploding as core collapse SNe, stars in the PISN mass range (i.e., $140\le m/\mathcal{M_{\odot}} \le 260$) exhibit only a very small neutron excess in their interiors (Heger \& Woosley 2002\nocite{heger02}). As a consequence, there is a pronounced odd-even effect, i.e., particularly low abundance ratios of odd-$Z$ elements to even-$Z$ elements, in the ejecta of PISNe. Furthermore, due to the lack of excess neutrons in addition to less-rapid expansion timescales during the explosion, the rapid n-capture process presumably does not take place in PISNe (Heger \& Woosley 2002\nocite{heger02}; Umeda \& Nomoto 2002\nocite{un02}). These and other chemical characteristics observable in low-metallicity stars, such as low values for $\mathrm{Fe}/\mathrm{Ca}$, could, in principle, be used to identify a possible pre-enrichment by PISNe.  However, despite considerable efforts made to identify and analyze the metal-poor tail of the Galactic halo MDF (see the in-depth review by Beers \& Christlieb 2005\nocite{bc05}), no observational evidence for a population of very massive stars exploding as PISNe has yet been found. It may appear, therefore, that only a negligible fraction of metal-free stars had masses in excess of $100~\mathcal{M_{\odot}}$. We will here argue, however, that this apparent absence of PISN progenitors might arise from a subtle observational selection effect. 

\begin{figure*}[t]
\resizebox{\hsize}{!}{\includegraphics{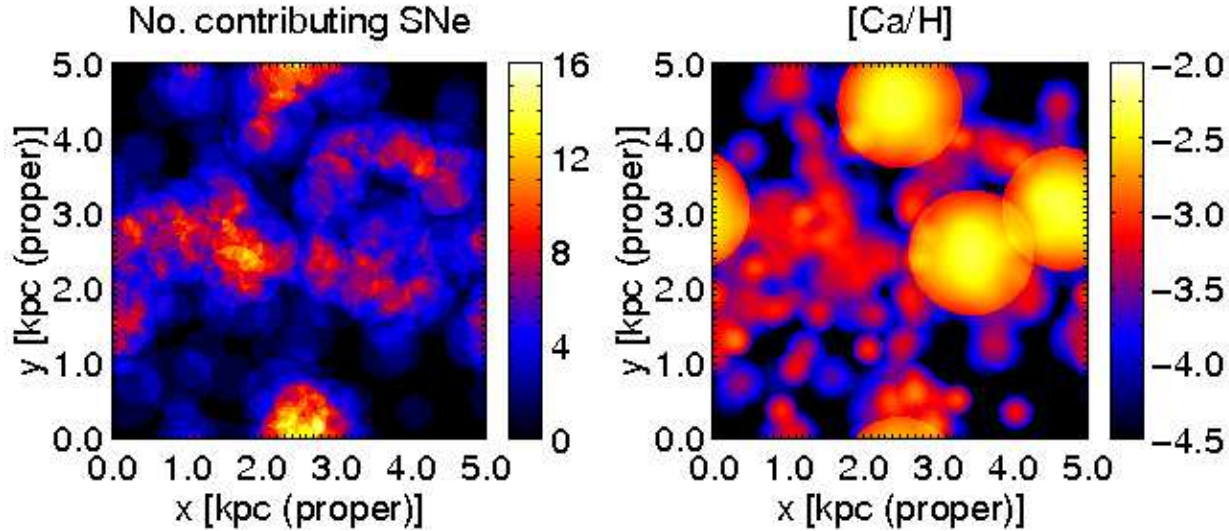}}
\caption{Illustration of the inhomogeneous chemical enrichment model. SNe of different types enrich space during the assembly of a $10^8~\mathcal{M_{\odot}}$ halo. {\it Left panel:} The number of contributing SNe in the interstellar medium at some early time. {\it Right panel:} The corresponding Ca abundance. Although the ejecta of PISNe are spread over very large volumes, the Ca abundance in regions enriched by a single PISN is significantly higher than in regions enriched even by numerous core collapse SNe.}
\label{illustr}
\end{figure*}

Classical, homogeneous chemical evolution models (see e.g., Prantzos 2005\nocite{prantzos05}; Ballero et al. 2006\nocite{ballero06}) predict that signatures of primordial PISNe, if they ever existed, by necessity must dominate the chemical abundance patterns found in the most metal-poor stars for two reasons: the very short lifetimes of the PISN progenitors and the assumption that SN ejecta are mixed instantaneously within the entire system.  Evidently, such calculations would suggest that all true second generation stars, i.e., stars enriched by a single primordial SN, will have the lowest metallicities of the observed metal-poor stars.  This, however, may not be the case.  As we will show, the chemical signature of primordial PISNe may only be found in stars of somewhat higher metallicity, possibly explaining why searches for the most metal-poor stars have not turned up any stars showing the PISN signature.

Although atomic diffusion in general is boosted by turbulence, giving rise to turbulent diffusion, the mixing of mass in the ISM is not instantaneous.  This fact becomes increasingly important in lower metallicity systems, as it implies that there should exist coeval regions in the ISM enriched by different types of SNe with distinctly different abundance patterns.  In particular, the amount of, e.g., calcium produced in a PISN can be several orders of magnitude higher than what is synthesized in a normal core collapse SN, with ratios such as $\mathrm{Fe}/\mathrm{Ca}$ that are only $\sim 1/10$ of those in  core collapse SNe.  A second generation star formed out of gas enriched by a single primordial PISN may therefore have a metal content which is atypical {\it and} significantly higher than a star formed from gas enriched by one, or even several core collapse SNe (see Bromm et al. 2003\nocite{byh03}; Greif et al. 2007\nocite{greif07}).  This effect is illustrated in Figure \ref{illustr} which shows, in a statistical sense, the spatial variation in the level of chemical enrichment that may be expected in the early universe.  Also shown is the corresponding variation in the number of SNe contributing to the enrichment.  Notice the poor correlation in certain regions between this number and the actual level of enrichment.  As a result of the non-instantaneous mixing of metals, which, by construction, is not handled in homogeneous chemical evolution models, the chemical relics of the very first stars in the universe (i.e., the primordial PISNe) could predominantly be ``hidden'' in a population of relatively metal-rich stars.

Since we are unable to follow the hierarchical build-up of the Milky Way halo in detail, we shall focus on the chemical enrichment of individual so-called `atomic cooling halos' of mass $\sim 10^8~\mathcal{M_{\odot}}$, which are able to cool due to lines of atomic hydrogen. These halos are probably massive enough to harbor continuous star formation while small enough to be dispersed during the subsequent assembly of the Galactic halo, either one-by-one or by first being incorporated into larger halos (see Robertson et al. 2005\nocite{robertson05}). Likely, a significant fraction of the metal-poor stars in the Galactic halo originate from such atomic cooling halos, in which the first SN-enriched gas may have re-collapsed and formed Pop~II stars (e.g., Greif et al. 2007\nocite{greif07}), including the second generation stars that we seek to identify. In accordance with this picture, Scannapieco et al. (2006)\nocite{scannapieco06} demonstrated that second generation stars should be expected over a wide range of Galactocentric radii. We shall assume that the star formation history of the atomic cooling halos begins with the first stars formed in collapsed minihalos. This picture is in agreement with hierarchical structure formation, where smaller systems merge to form larger ones. In the following, star formation will be divided into two modes: Pop~III star formation, which only occurs in metal-free gas, and Pop~II star formation, occurring in gas already enriched with metals.

\section{Modeling Early Chemical Enrichment}\label{ece}
In order to discern the chemical signature of the first massive Pop~III stars, we have constructed a model for how the early universe was enriched with heavy elements ejected by the first SNe. Our model accounts, self-consistently, for the formation of both Pop~III and Pop~II stars, and allows us to study in some detail how SNe from each of these populations contributed to the abundances that can be observed today in Galactic metal-poor stars. In this section, we describe our chemical evolution model, emphasizing how it incorporates the distinct characteristics of Pop~III star formation in a cosmological setting.

\subsection{The Basic Model}\label{model}
Although the details of the inhomogeneous chemical enrichment model are described elsewhere (see Karlsson 2005\nocite{karlsson05}, 2006\nocite{karlsson06}; Karlsson \& Gustafsson 2005\nocite{kg05}), we will provide a general outline and discuss relevant improvements below. The model is based on the mixing volume picture, where the metal-rich ejecta of individual SNe are spread within finite but continuously growing volumes $V_{\mathrm{mix}}(t)$. In this picture, the mixing volume growth is modeled as a random-walk process, predominantly driven by turbulent gas motions. At any given time $t$, we may thus define the size of $V_{\mathrm{mix}}$ as

\begin{equation} 
V_{\mathrm{mix}}(t) = \frac{4\pi}{3}(6D_{\mathrm{t}}t+\sigma_E)^{3/2},
\label{vmix}
\end{equation}

\noindent
where $D_{\mathrm{t}}=\langle v_{\mathrm{t}}\rangle l_{\mathrm{t}}/3$ is the turbulent diffusion coefficient and $\sigma_E$ is the minimal mixing area, introduced to account for the initial expansion of the ejected material, before it merges with the ambient medium. Furthermore, in accordance with the diffusion approximation, the concentration of metals within $V_{\mathrm{mix}}$ is assumed to be non-uniform, following a Gaussian density profile. This was not taken into account in earlier versions of the model. The turbulent diffusion coefficient is estimated to be $D_{\mathrm{t}}\simeq 3.3\times 10^{-4}$ kpc$^2$ Myr$^{-1}$, assuming a typical turbulent velocity of $\langle v_{\mathrm{t}}\rangle \simeq v_{\mathrm{vir}}\simeq 10$ km s$^{-1}$, where $v_{\mathrm{vir}}$ is the virial velocity of  $\sim 10^8~\mathcal{M_{\odot}}$ atomic cooling halos, and a turbulent correlation length of $l_{\mathrm{t}}=100$ pc, a typical value for the virial radius of a minihalo, on which scale mergers may be expected to drive turbulent mixing.

During the early expansion of the SN remnant itself, no star formation is expected to occur within $V_{\mathrm{mix}}$.  This is taken into account by assuming an initial, instantaneous growth of $V_{\mathrm{mix}}$ set by $\sigma_E$.  The value of $\sigma_E$ depends on how much mass of gas is swept up by the remnant, which is chiefly determined by the explosion energy of the SN and only very weakly dependent on the density and metallicity of the ambient medium (Cioffi et al. 1988\nocite{cioffi88}).  For core collapse SNe, this initial dilution mass is estimated to be $M_{\mathrm{dil}}\simeq 10^5~\mathcal{M_{\odot}}$ assuming an explosion energy of $E=10^{51}$~ergs, an average ISM metallicity of $Z/Z_{\odot}=10^{-3}$, and a density of $n\simeq 0.1$ cm$^{-3}$ (the weak dependences on $Z$ and $n$ are neglected, see Ryan et al. 1996\nocite{rnb96} and references therein), while for PISNe, $M_{\mathrm{dil}}\simeq 5\times 10^5(E/10^{51}~\mathrm{ergs})^{0.6}~\mathcal{M_{\odot}}$ at $z\sim 20$ (Bromm \& Loeb 2003\nocite{bromm03}). Depending on the PISN explosion energy, the associated initial dilution mass may be nearly two orders of magnitude larger than the corresponding dilution mass for core collapse SNe. Clearly, the inclusion of this effect is important as it may strongly influence the abundance of chemical elements in regions where subsequent stars eventually are allowed to form (see Fig. \ref{illustr}).

The present model is modified to handle the statistics of the metal-free Pop~III and the metal-enriched Pop~II separately.  Pop~II star formation is assumed to be spatially uncorrelated.  This means that the probability, $w_{\mathrm{I\!I}}(k,t)$, of finding a region in space enriched by the ejecta of $k$ Pop~II SNe at time $t$ can be approximated by the Poisson distribution, i.e., 

\begin{equation}
w_{\mathrm{I\!I}}(k,t) = \mathrm{Po}(k,\mu_{\mathrm{I\!I}}(t))=e^{-\mu_{\mathrm{I\!I}}(t)}\mu_{\mathrm{I\!I}}(t)^k/k!,
\label{wii}
\end{equation}

\noindent
where $\mu_{\mathrm{I\!I}}(t)$ denotes the space-averaged number of Pop~II SNe contributing to the enrichment at time $t$.  For Pop~III star formation, however, the situation is somewhat different.  If Pop~III stars would have been randomly distributed in those areas where star formation occurred, i.e., in the higher density filaments, the probability of finding a region in such a filament enriched by a given number $k$ of Pop~III SNe would have been described by equation (\ref{wii}).  However, the fact that no metal-free star is allowed to form in a region previously enriched by metals (but see Wyithe \& Cen 2007\nocite{wc07}; Cen \& Riquelme 2007\nocite{cen07}) introduces a type of anti-clustering effect.  The fraction of enriched gas will thus be larger than that described by the Poisson statistics, for any given value of $\mu_{\mathrm{I\!I\!I}}(t)>0$, the space-averaged number of contributing Pop~III stars at time $t$ (see \S \ref{starformation}).  To approximately account for this ``no overlapping''-effect, the Poisson distribution is empirically modified by a Gaussian. Thus, the probability $w_{\mathrm{I\!I\!I}}$ of finding a region enriched by $k$ Pop~III stars is given by 

\begin{equation}
w_{\mathrm{I\!I\!I}}(k,t) = c_k e^{-(\mu_{\mathrm{I\!I\!I}}(t)-k)^2/4} \times \mathrm{Po}(k,\mu_{\mathrm{I\!I\!I}}(t)), 
\label{wiii}
\end{equation} 

\noindent
where $c_k$ is a normalization factor.  In contrast to $w_{\mathrm{I\!I}}$, the probability density function described by \mbox{equation (\ref{wiii})} is more narrowly peaked towards $\mu_{\mathrm{I\!I\!I}}=k$. In particular, $w_{\mathrm{I\!I\!I}}$ predicts a faster-than-exponential decrease of the fraction of primordial gas available to Pop~III star formation, as a function of $\mu_{\mathrm{I\!I\!I}}$. With time, $w_{\mathrm{I\!I\!I}}(k=0,t)$ will still decrease slower than predicted by the corresponding Poisson equation since the Pop~III SN rate, in fact, is proportional to the available fraction of primordial gas, as will be further discussed in \S \ref{starformation}.

Given the expressions in equation (\ref{wii}) and (\ref{wiii}), we may calculate the probability $f_{\mathrm{I\!I\!I}}(k',k)$ that a low-mass star (here defined to be a star which has survived to the present, i.e., with a mass $\lesssim 0.8~\mathcal{M_{\odot}}$) is formed out of gas enriched by a total number $k$ of SNe (i.e., Pop~III $+$ Pop~II), $k'$ of which are Pop~III SNe. This probability is given by the integral 

\begin{equation}
f_{\mathrm{I\!I\!I}}(k',k) = \tilde{c}_k\int\limits_{0}^{\tau_{\mathrm{H}}} a_{\star}(t)w_{\mathrm{I\!I\!I}}(k',t)w_{\mathrm{I\!I}}(k-k',t)u_{\mathrm{I\!I}}(t)\mathrm{d}t.
\label{fkprimk}
\end{equation}

\noindent
Here, $u_{\mathrm{I\!I}}$ denotes the formation rate of Pop~II stars expressed in terms of the resulting SN rate (see \S \ref{starformation}), while $a_{\star}(t)$ is the fraction of still surviving stars in a stellar generation formed at time $t$ and $\tilde{c}_k$ is a normalization factor (note that $\tilde{c}_k \not=c_k$). The age of the system is set to $\tau_{\mathrm{H}}\simeq 10^{10}$ yr to account for late-time star formation in the Milky Way halo. Equation (\ref{fkprimk}) is used to predict the Pop~III contribution to the amount of metals in the gas from which the low-mass stars were formed.  Following Karlsson (2005\nocite{karlsson05}, 2006\nocite{karlsson06}), we make use of Equations (\ref{vmix})$-$(\ref{fkprimk}) to calculate the predicted relative number density of still surviving ($m\lesssim 0.8~\mathcal{M_{\odot}}$) metal-poor stars with any given chemical composition. These density functions are then analyzed and compared to observations. In the remainder of this section, we will define and discuss the other model parameters.

\subsection{Density Evolution}\label{density}
Due to the inability of our chemical enrichment model to account for spatial variations, except in a statistical sense, we will adopt the average density evolution $n=n(t)$ of a spherically collapsing $10^8~\mathcal{M_{\odot}}$ halo embedded in an expanding universe.  We assume that the density of the halo follows the top-hat model for its collapse and virialization (e.g., Padmanabhan 2002\nocite{padman02}; Bromm et al. 2002\nocite{bromm02}). After virialization at $z_{\mathrm{vir}}\simeq 10$, a constant density of \mbox{$n=n_{\mathrm{vir}}=0.1~\mathrm{cm}^{-3}$} is assumed, where $n_{\mathrm{vir}}$ denotes the average density at virialization.

\subsection{Initial Mass Functions and Stellar Yields}\label{imfs}
The relative distribution by mass for metal-free and extremely metal-poor stars is largely unknown. This is especially true for the high-mass end of the IMF since all metal-free, massive stars in our part of the universe are long gone. We must therefore resort to theoretical predictions. We shall assume that the IMF of the metal-enriched Pop~II follows that of the solar neighborhood, which can be approximated with a broken power-law, i.e., \mbox{$\mathrm{d}N/\mathrm{d}m \equiv \phi(m) \propto m^{\alpha}$}, with a Salpeter-like exponent of $\alpha=-2.3$ above $m=0.5~\mathcal{M_{\odot}}$ and a shallower slope of $\alpha=-1.3$ below this mass (Kroupa 2001\nocite{kroupa01}). Very massive stars exploding as PISNe are assumed to form exclusively in metal-free gas and are therefore not formed in the Pop~II mode (but see Langer et al. 2007\nocite{langer07}).  Conversely, low-mass stars are only allowed to form in the Pop~II mode. In addition, following Bromm \& Loeb (2003\nocite{bromm03}), low-mass star formation is restricted to occur in regions with efficient cooling, i.e., where the abundance of carbon and/or oxygen is high enough. To include this effect, we adopt the metallicity criterion $D_{\mathrm{trans}}$, introduced by Frebel et al. (2007\nocite{frebel07}), which takes into account the simultaneous cooling by O~{\sc i} and C~{\sc ii}. Hence, low-mass stars are only able to form in gas with C and O abundances such that 

\begin{displaymath}
D_{\mathrm{trans}}=\log_{10}\left(10^{[\mathrm{C}/\mathrm{H}]}+0.3\times 10^{[\mathrm{O}/\mathrm{H}]}\right)>-3.5 \mbox{\ .} 
\end{displaymath}

The special initial conditions in the first collapsing minihalos presumably favored the formation of a population of very massive stars.  Although these conditions are not always met in metal-free gas in general (see, e.g., Johnson \& Bromm 2006\nocite{jb06}; Padoan et al. 2007\nocite{padoan07}), the top-heaviness of the primordial IMF seems to persist.  In particular, the apparent lack of metal-free stars in the present Galactic halo indicates that primordial low-mass star formation was highly suppressed (e.g., Karlsson 2005\nocite{karlsson05}; Tumlinson 2006\nocite{tumlinson06}).  For the Pop~III IMF, we shall assume that stars below $10~\mathcal{M_{\odot}}$ were not able to form. Above this limit, stars in the mass range of core collapse SNe are assumed to be distributed in the same way as for Pop~II, i.e., as a power-law with a slope of $\alpha=-2.3$. Note that although Padoan et al. (2007\nocite{padoan07}) derive a primordial/metal-poor IMF with a slope which asymptotically approaches $\alpha = -3.5$ at the high-mass end, the slope in the core collapse range is significantly shallower.

As regards the shape of the unknown, metal-free very-high-mass IMF, we take a conservative approach and assume that stars in the mass range of PISNe are all formed with equal probability, independent of their individual masses. As a measure of the fraction of very massive Pop~III stars, we introduce the parameter $\beta$. This parameter is defined as the ratio between the number of PISNe and the total number of exploding stars in a Pop~III generation. With this definition, $\beta$ is a lower limit to the original fraction of very massive Pop~III stars, as a fraction of these stars will become black holes by direct collapse instead of exploding as PISNe (e.g., Heger et al. 2003\nocite{heger03}).

In the following, $\beta$ will be considered as a free model parameter, and, as we shall see, it is also one of the parameters to which the result is most sensitive. If $\beta = 1$, all Pop~III stars were formed as very massive stars ending their lives as PISNe, while if $\beta = 0$, no PISN ever occurred in the early universe. In later sections, we shall constrain $\beta$ by comparing observations with model results. It is, however, possible to estimate the plausible range for $\beta$ from a theoretical point of view. By extrapolating the Galactic IMF (Kroupa 2001\nocite{kroupa01}) to very high masses, $\beta$ is estimated to be $0.032$. Alternatively, using the more realistic primordial IMF by Padoan et al. (2007\nocite{padoan07}), a $\beta$ of $0.017$ is obtained. Greif \& Bromm (2006\nocite{gb06}) estimated the mass fraction of metal-free stars that went into very massive stars by considering the relative number of minihalos and atomic-cooling halos in the early universe. They derived a value of $5-10\%$ by mass, which translates into a $\beta \simeq 0.005-0.01$, based on the idea that the PISNe and core collapse SNe were formed in different types of halos, the former in minihalos and the latter in atomic halos.

We consider here only those chemical elements which are predominantly synthesized in massive stars, e.g., calcium and iron. We will not take into account the limited enrichment by intermediate-mass stars and thermonuclear (Type Ia) SNe, whose main contributions to the chemical enrichment occur at a considerably later stage in the history of the Milky Way. The ejection of newly synthesized material will be restricted to two mass ranges: that of core collapse SNe and that of PISNe (Heger \& Woosley 2002\nocite{heger02}). The stellar yields of core collapse SNe, with progenitor masses in the range $13\le m/\mathcal{M_{\odot}}\le 40$, are taken from the new calculations by Nomoto et al. (2006\nocite{nomoto06}), while the yields of the PISNe ($140\le m/\mathcal{M_{\odot}}\le 260$) are taken from Heger \& Woosley (2002)\nocite{heger02}. Core collapse SNe with progenitor masses $\lesssim 13~\mathcal{M_{\odot}}$ are believed to synthesize very few heavy elements (see, e.g., Nomoto 1987\nocite{nomoto87}; Mayle \& Wilson 1988\nocite{mw88}) and are neglected here. As an estimate of the uncertainty in the results due to uncertainties in the nucleosynthesis calculations (see \S \ref{uncertainties}), we also ran a simulation using the yields of PISNe by Umeda \& Nomoto (2002\nocite{un02}).

\subsection{Star Formation Rates}\label{starformation}
As discussed above, the metal-free, Pop~III mode of star formation likely differed from that of Pop~II, occurring in metal enriched gas. We will account for this distinction by introducing one SFR for the Pop~III mode and a different one for the Pop~II mode. Below, these SFRs will be displayed in terms of their respective SN rate in units of kpc$^{-3}$ Myr$^{-1}$. Stars not ejecting any metals, e.g., stars subject to direct black hole formation, are not included in these rates.  Again, low-mass star formation is considered to occur in Pop~II only, with the low-mass star formation rate closely following that of Pop~II SNe.

\begin{figure}[t]
\resizebox{\hsize}{!}{\includegraphics{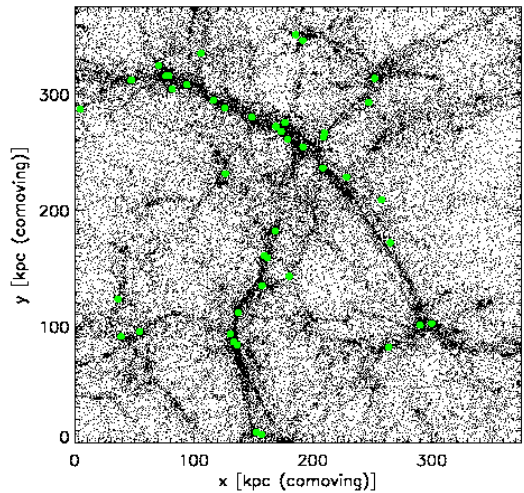}}
\caption{Sites of Pop~III star formation in our SPH simulation. Shown here in projection is the density field around the $\sim$ 10$^8$ M$_{\odot}$ system in our cosmological box at redshift $z$ $\sim$ 10. The green dots indicate the sites where the primordial gas collapses into minihalos and so allows the formation of Pop~III stars. We estimate the resulting star formation rate of Pop~III stars by assuming that a single Pop~III star forms in each of these minihalos.}
\label{sf}
\end{figure}

We carried out a smoothed particle hydrodynamics (SPH) simulation of the assembly of a $10^8~\mathcal{M_{\odot}}$ halo, including star formation without feedback, in order to estimate the star formation rate for Pop~III stars forming in minihalos which are later incorporated into the larger halo.  The simulation is set up similarly to those conducted in Johnson et al. (2007\nocite{jgb07}), employing a cosmological box of comoving length $660$~kpc. We track the sites where the primordial gas is able to collapse to form Pop~III stars in this simulation (Fig. \ref{sf}), and we thereby estimate the Pop~III star formation rate within the volume being assembled into the $10^8~\mathcal{M_{\odot}}$ halo, assuming that a single Pop~III star forms in each of the collapsing minihalos (e.g., Yoshida et al. 2006\nocite{yoshida06}). In the simulation, the first star forms at a redshift of $z\simeq 23$. We find that the Pop~III star formation rate, here expressed as the resulting SN rate, is $\sim$ 0.04 kpc$^{-3}$ Myr$^{-1}$ (physical) within the region which is finally virialized in this halo.  We thus use this value for the normalization of the Pop~III star formation rate.  We note that the neglect of local radiative and mechanical feedback in this simulation is not likely to greatly affect the value that we find for the star formation rate (see Susa \& Umemura 2006\nocite{susa06}; Ahn \& Shapiro 2007\nocite{ahn07}; Johnson et al. 2007\nocite{jgb07}; Greif et al. 2007\nocite{greif07}; but see Whalen et al. 2007\nocite{whalen07}).

\begin{figure}[t]
\resizebox{\hsize}{!}{\includegraphics{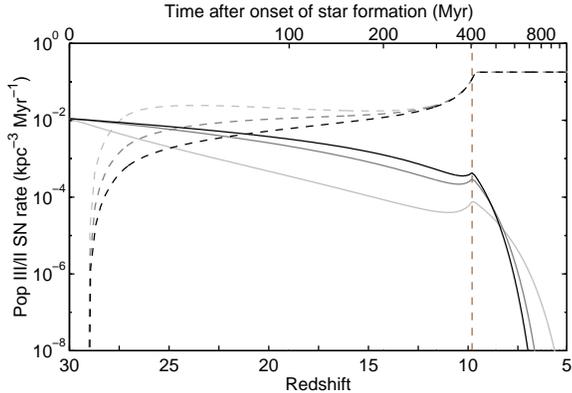}}
\caption{Adopted Pop~III (solid lines) and Pop~II (dashed lines) SN rates for $\beta=1,~0.1$, and $0.01$ (darker lines correspond to smaller $\beta$). The dashed, vertical (brown) line marks the virialization redshift of the $10^8~\mathcal{M_{\odot}}$ halo. In general, as an increasing volume of space becomes metal enriched, the Pop~III rates decrease while the Pop~II rates increase. The bump in the Pop~III rates at $z=10$ is due to a sharp rise in density prior to virialization. The small delay in the rise of the Pop~II rates is due to the finite lifetimes of Pop~II core collapse SNe.
}
\label{pop3}
\end{figure}  

The star formation rate is assumed to be proportional to the gas density $n(t)$. Hence,

\begin{equation}
u_{\mathrm{I\!I\!I}}(t) = u_{\mathrm{I\!I\!I,0}}\times Q_{\mathrm{p}}(t) \left(\frac{n(t)}{0.1~\mathrm{cm}^{-3}}\right),
\label{pop3eq}
\end{equation}

\noindent
and

\begin{equation}
u_{\mathrm{I\!I}}(t) = u_{\mathrm{I\!I,0}}\times \left(1-Q_{\mathrm{p}}(t)\right)\left(\frac{n(t)}{0.1~\mathrm{cm}^{-3}}\right),
\label{pop2eq}
\end{equation}

\noindent
where $u_{\mathrm{I\!I\!I}}$ and $u_{\mathrm{I\!I}}$ denote the Pop~III and Pop~II SN rate, respectively (Fig. \ref{pop3}), and $Q_{\mathrm{p}}$ denotes the volume filling factor of primordial gas. The Pop~III rate is normalized to $u_{\mathrm{I\!I\!I,0}}=0.04$ kpc$^{-3}$ Myr$^{-1}$ (see above). For metal enriched gas, the star formation efficiency is assumed to be $10$ times higher. The Pop~II rate is therefore normalized to $u_{\mathrm{I\!I,0}}=0.4$ kpc$^{-3}$ Myr$^{-1}$ which is close to the average SN rate in the galactic halo, derived from chemodynamical modeling (Samland et al. 1997\nocite{samland97}). The fraction of stars predominantly enriched by PISNe will not be very sensitive to these normalizations (see Table \ref{parameters}). We assume that star formation begins at $z=30$, to account for the rare high-$\sigma$ peaks in the dark matter distribution.

In the above expressions, the presence of the factor $Q_{\mathrm{p}}$ is important as it regulates the amount of gas available for both the Pop~III and the Pop~II mode of star formation. It is controlled by the Pop~III SFR via Equation (\ref{wiii}). Hence, for $k=0$, 

\begin{equation}
Q_{\mathrm{p}}(t) \equiv w_{\mathrm{I\!I\!I}}(0,t) = e^{-\mu_{\mathrm{I\!I\!I}}(t)-\mu_{\mathrm{I\!I\!I}}(t)^2/4}.
\label{qz}
\end{equation}

\noindent
Since the Pop~III SN rate depends on the available amount of primordial gas which, in turn, is determined by the rate of metal enrichment, $u_{\mathrm{I\!I\!I}}$ and $Q_{\mathrm{p}}$ are coupled via the expression for $\mu_{\mathrm{I\!I\!I}}$:

\begin{eqnarray}
\mu_{\mathrm{I\!I\!I}}(t) & = & \beta \int\limits_{0}^{t} V_{\mathrm{mix}}^{\gamma\!\gamma}(t-t') u_{\mathrm{I\!I\!I}}(t')\mathrm{d}t'\nonumber\\
& + & (1-\beta) \int\limits_{0}^{t} V_{\mathrm{mix}}^{\mathrm{cc}}(t-t') u_{\mathrm{I\!I\!I}}(t')\mathrm{d}t'.
\label{muiii}
\end{eqnarray}

\noindent
In this expression, the contribution from PISNe and core collapse SNe are distinct since $\sigma_E$, which controls the initial expansion of the ejecta, differs between the two types of SNe (see above). Here, $V_{\mathrm{mix}}^{\gamma\!\gamma}$ and $V_{\mathrm{mix}}^{\mathrm{cc}}$ denote the mixing volumes for PISNe and core collapse SNe, respectively, while $\beta$, defined in \S \ref{imfs}, denotes the fraction of PISNe in Pop~III.

\section{Results}\label{results}
In this section, we shall discuss our results for the chemical enrichment during the assembly of atomic-cooling halos and compare them with recent observations.  We would like to emphasize that not all stars belonging to the Milky Way halo were formed in atomic cooling halos, in particular the stars at the metal-rich end (cf. Robertson et al. 2005\nocite{robertson05}).  However, a significant fraction of the stars below $[\mathrm {Ca}/\mathrm{H}]=-2$ and a majority of the second generation stars presumably originated from this type of halo.\footnote{$[\mathrm{A}/\mathrm{B}]=\log_{10}(N_A/N_B)_{\star}-\log_{10}(N_A/N_B)_{\odot}$, where $N_X$ is the number density of element $X$.} If so, they are likely to be found over the entire Galactic halo (Scannapieco et al. 2006\nocite{scannapieco06}). With this in mind, we will make no distinction in origin between our simulated stars and the field stars observed in the Galactic halo. 

\begin{figure}[t]
\resizebox{\hsize}{!}{\includegraphics{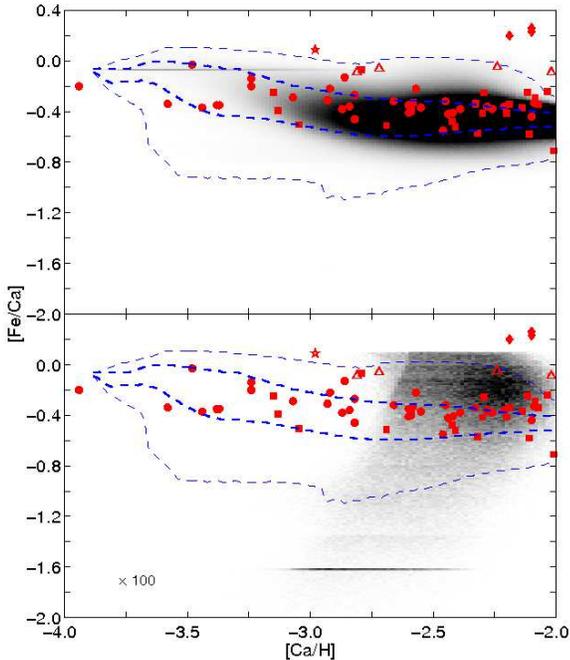}}
\caption{\scriptsize{Predicted distributions (shaded areas) of low-mass stars in the $[\mathrm{Fe}/\mathrm{Ca}]-[\mathrm{Ca}/\mathrm{H}]$ plane for $\beta=0.1$. {\it Upper panel:} The full distribution. Filled (red) circles are observations of giants in the Galactic halo by Cayrel et al. (2004), filled (red) squares are observations of dwarfs by Cohen et al. (2004), filled (red) diamonds are three field stars depleted in n-capture elements (Ivans et al. 2003), while open (red) triangles are observations of giants in dwarf spheroidal galaxies (Shetrone et al. 2001), and the (red) star is Draco 119 (Fulbright et al. 2004). The dashed, thick and thin (blue) lines indicate, respectively, the $1\sigma$ (innermost $68.3\%$) and $3\sigma$ (innermost $99.7\%$) cosmic scatter in the simulation. {\it Lower panel:} The partial distribution of simulated stars for which $>90\%$ of the total atmospheric Ca abundance is synthesized in PISNe. This distribution is multiplied by a factor of $100$. The observations and the dispersion of the full distribution are plotted for comparison. Note, the mildly enhanced density of simulated stars at $[\mathrm{Fe}/\mathrm{Ca}]\simeq-1.6$ originates from a nearly constant Fe/Ca yield ratio in the mass range $160\lesssim m/\mathcal{M_{\odot}} \lesssim 170$.}}   
\label{fecah}
\end{figure}
\nocite{cayrel04}
\nocite{cohen04}
\nocite{ivans03}
\nocite{shetrone01}
\nocite{fulbright04}

Instead of iron, we choose calcium as the reference element for the metallicity of the stars.  We do so for two reasons.  First, the strong $\mathrm{Ca}$ {\sc ii} K resonance line is commonly used as metallicity indicator (Beers et al. 1990\nocite{beers90}) in surveys of metal-poor stars, such as the HK survey (Beers et al. 1992\nocite{beers92}) and the Hamburg/ESO survey (Wisotzki et al. 2000\nocite{wisotzki00}).  The much weaker Fe lines will remain undetected in low-quality survey spectra, although they can be detected by taking high-resolution spectra.  This is of importance since the vast majority of the most metal-poor stars known thus were originally recognized as being metal-poor on the basis of their Ca abundance, not their Fe abundance.   The second reason we choose calcium is that the iron yield is, to a larger extent than calcium, affected by the unknown amount of fall-back in core collapse SNe and it varies by more than $3$ orders of magnitude over the PISN mass range.  The $\mathrm{Ca}$ yield, on the other hand, varies by merely $1$ order of magnitude (Heger \& Woosley 2002\nocite{heger02}; Umeda \& Nomoto 2002\nocite{un02}). For the employed observational data, either existing $\mathrm{Ca}$ abundances were used or, in case of the observed Galactic halo MDF, an offset of $[\mathrm{Ca}/\mathrm{Fe}]=+0.4$ dex was applied to convert from $\mathrm{Fe}$ abundances to $\mathrm{Ca}$ abundances.  This offset is based on the average Ca excess (Pagel \& Tautvai\v{s}ien\.{e} 1995\nocite{pagel95}) observed in Galactic halo stars (see e.g., Cayrel et al. 2004\nocite{cayrel04}), although there are outlier stars in the lowest metallicity range which deviate from the mean $[\mathrm{Ca}/\mathrm{Fe}]=+0.4$.  Apart from the offset, however, the shape of the Galactic halo MDF probably doesn't change much whether Fe or Ca is used as the metallicity indicator.

\subsection{Comparison with Observed Metal-Poor Stars}
In Figure \ref{fecah}, we show the probability density function of low-mass stars in the $[\mathrm{Fe}/\mathrm{Ca}]-[\mathrm{Ca}/\mathrm{H}]$ plane from a simulation in which $10\%$ by number (i.e. $\beta = 0.1$), or about $50\%$ by mass, of all Pop~III stars were allowed to explode as PISNe. The agreement between the model and the observations of Galactic field stars by Cayrel et al. (2004\nocite{cayrel04}) and Cohen et al. (2004\nocite{cohen04}) is satisfactory.  It should be noted that the full distribution displayed in the upper panel of Figure \ref{fecah} is effectively unaltered for any $\beta\lesssim 0.1$.

The surprisingly small star-to-star scatter and the lack of evolution in the scatter that is observed in various abundance ratios for well defined samples of extremely metal-poor Galactic halo stars (e.g., Cayrel et al. 2004\nocite{cayrel04}; Arnone et al. 2005\nocite{arnone05}; Barklem et al. 2005\nocite{barklem05}) is not fully understood.  Such a small scatter could be an indication of extremely short mixing timescales and a well mixed ISM already at very early times.  However, as evident from Figure \ref{fecah}, the small scatter is as well reproduced by our model over the entire metallicity regime. Despite the relatively slow mixing assumed in our model, the small dispersion is explained by a selection effect favoring contributions from SNe in a certain mass range ($13\le m/\mathcal{M_{\odot}} \lesssim 20$, in this particular case) for the most metal-poor stars and by the averaging of a large number of contributing SNe at higher metallicities (Karlsson \& Gustafsson 2005\nocite{kg05}). Ultimately depending on the stellar yields, a small star-to-star scatter can thus be realized without invoking unrealistically short mixing timescales in the early ISM.

\subsection{Characteristics of PISN-Dominated Stars}
In the lower panel of Figure \ref{fecah}, only those simulated low-mass stars are displayed for which $>90\%$ of the total atmospheric calcium abundance was synthesized in PISNe.  This regards the abundance originally present in the gas out of which the stars were formed and not the abundance at late-stage evolution, although the difference should be negligible for Ca.  Henceforth, such stars will be referred to as PISN-dominated stars.  From the results depicted in Figure \ref{fecah}, three conclusions regarding the chemical legacy of PISNe can immediately be drawn. First and foremost, the low-mass stars with a dominant contribution from PISNe, as reflected by the Ca abundance, are located at significantly higher $[\mathrm{Ca}/\mathrm{H}]$ than the most Ca deficient stars in the simulation, which are found at $[\mathrm{Ca}/\mathrm{H}]\simeq -4$.  The PISN-dominated stars are instead found in the range $-3 \lesssim [\mathrm{Ca}/\mathrm{H}]\lesssim -2$, with a maximum around $[\mathrm{Ca}/\mathrm{H}]\simeq -2.3$.  This prediction is at variance with the classical picture in which the chemical signatures of the first, very massive stars are anticipated to be found in the most metal-poor regime. This is exactly the effect illustrated in Figure \ref{illustr} and described in \S \ref{scenario}. Looking for the stars with the lowest abundances of Ca (or Fe) may thus not be the best way to find the relics of the first PISNe. 

\begin{figure}[t]
\resizebox{\hsize}{!}{\includegraphics{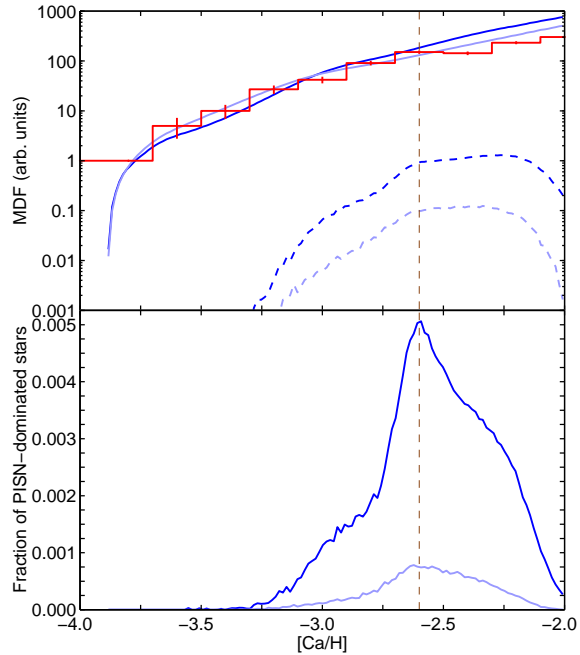}}
\caption{\small{Predicted metallicity distribution functions as measured by $[\mathrm{Ca}/\mathrm{H}]$ for $\beta=0.1$ (dark blue lines) and $\beta=0.01$ (light blue lines). {\it Upper panel:} The solid lines denote the total MDFs while the dashed lines denote the MDFs of PISN-dominated stars, for which $>90\%$ of the Ca originate from PISNe. The binned data (red line) are the observed (not completeness corrected) distribution of Galactic halo stars from the Hamburg/ESO survey by Beers et al. (2005), where the vertical lines indicate the ($1\sigma$) stochastic uncertainty. Note that these data are shifted $+0.4$ dex. {\it Lower panel:} The fraction of PISN-dominated stars. The majority of these stars fall above the upper limit of $[\mathrm{Ca}/\mathrm{H}]=-2.6$ (dashed brown line), the range where the Hamburg/ESO survey is incomplete.}}
\label{mdfs}
\end{figure}
\nocite{beers05}

Second, the fraction of PISN-dominated stars in the halo is predicted to be small.  As illustrated in the lower panel of Figure \ref{fecah}, the distribution of PISN-dominated stars has to be multiplied with a factor of $100$ in order to be detected on the same scale as the total distribution, shown in the upper panel. Even for a relatively large $\beta$ of $0.1$ (i.e., $\sim 50\%$ by mass), the total fraction of PISN-dominated stars below $[\mathrm{Ca}/\mathrm{H}]=-2$ is merely $\sim 2\times10^{-3}$, or about one star in $500$. For lower $\beta$, this fraction decreases further (see \S \ref{obsconstr}).

Third, as noted in \S \ref{scenario}, the chemical signature of PISNe distinctly differs from that of the less massive core-collapse SNe. This provides an opportunity to identify outliers, at relatively high metallicities, as possible PISN-enriched, second generation stars. As shown in Figure \ref{fecah} (lower panel), PISN-dominated stars are predicted to be found both above and below the IMF-averaged value of $[\mathrm{Fe}/\mathrm{Ca}]\simeq -0.4$. Noticeably, none of the stars in the samples by Cayrel et al. (2004\nocite{cayrel04}) and Cohen et al. (2004\nocite{cohen04}) show a large enough deviation from the mean $[\mathrm{Fe}/\mathrm{Ca}]$, and other abundance ratios, to be classified as potential PISN-enriched stars. However, the four stars belonging to the Draco, Ursa Minor, and Sextans dwarf spheroidal (dSph) galaxies have a relatively high $[\mathrm{Fe}/\mathrm{Ca}]\simeq -0.1$. These gas-poor satellite galaxies primarily consist of old and metal-poor stars and were presumably formed at an early stage, as a result of the collapse of dark-matter halos of $\lesssim 10^9~\mathcal{M_{\odot}}$.

Could these dSph stars be second generation, PISN-dominated stars? This seems unlikely. The low $\alpha$-element/iron ratios are generally attributed to Type Ia SNe in connection to lower SFRs in these systems (e.g., Matteucci 2003\nocite{matteucci03}), and, while no contribution of s-process elements by asymptotic giant branch stars is detected in dSph stars below $[\mathrm{Fe}/\mathrm{H}]\simeq -1.8$ the high Ba$/$Y and low Y$/$Eu ratios (as well as low $\alpha/\mathrm{Fe}$) could be explained by the absence of elements synthesized in the $\alpha$-process (Venn et al. 2004\nocite{venn04} and references therein), suggested to occur in hypernovae (Nakamura et al. 2001\nocite{nakamura01}). A few other, curious outliers such as Draco $119$ (Fulbright et al. 2004\nocite{fulbright04}) and BD$~+80^{\circ}~245$, G$4$-$36$, and CS $22966$-$043$ studied by Ivans et al. (2003\nocite{ivans03}), will be further discussed in \S \ref{discussion}.

\begin{figure}[t]
\resizebox{\hsize}{!}{\includegraphics{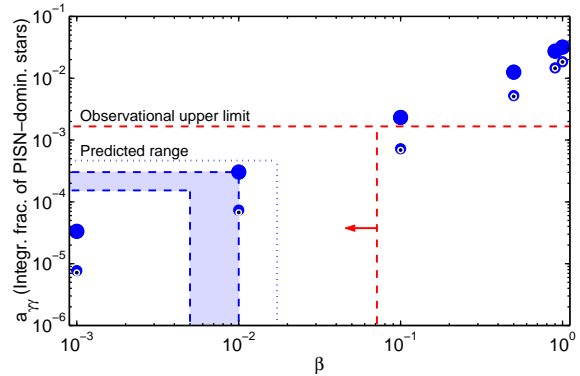}}
\caption{\small{The predicted integrated (total) fraction of PISN-dominated stars below $[\mathrm{Ca}/\mathrm{H}]=-2$ as a function of $\beta$, corresponding to $>90\%$ (big blue dots), $>99\%$ (medium sized, blue dots), and $>99.9\%$ (small, dark blue dots) PISN-enrichment.  The dashed (red) lines indicate the observational upper limit of $\beta$, assuming that none of the $\sim 600$ Galactic halo stars with $[\mathrm{Ca}/\mathrm{H}]\le -2$ for which high-resolution spectroscopy is available show any signature of PISNe (N. Christlieb, priv. comm.).  The dotted (blue) line and shaded (blue) area denote the predicted range of $a_{\gamma\!\gamma}$ anticipated from the data of Padoan et al. (2007) and Greif \& Bromm (2006), respectively.  If nothing else stated, it is assumed that the PISN signature is detectable in Galactic halo stars with $>90\%$ PISN-enrichment.}}
\label{apisn}
\end{figure}
\nocite{padoan07}
\nocite{gb06}

In the simulations, the distribution of stars with a dominant contribution from PISNe peaks around $[\mathrm{Ca}/\mathrm{H}]\simeq -2.5$, or even higher (see Fig. \ref{fecah}, lower panel and Fig. \ref{mdfs}, upper panel).  In fact, the majority of PISN-dominated stars are predicted to be found above \mbox{$[\mathrm{Ca}/\mathrm{H}]=-2.6$} (corresponding to \mbox{$[\mathrm{Fe}/\mathrm{H}]=-3$}).  This is the upper limit for which the Hamburg/ESO survey is considered to be complete (see Fig. \ref{mdfs}, upper panel). A significant fraction of the second generation stars may therefore have a chance to remain undetected in the Hamburg/ESO survey, not because they are too Ca-poor but because they are too Ca-rich.  Note the deficit above $[\mathrm{Ca}/\mathrm{H}]=-2.6$ in the observed MDF which, to a large extent, has to be attributed to this intended observational bias against metal-rich stars (Fig. \ref{mdfs}, upper panel).  With this taken into account, the observed and predicted MDFs, in particular the one with $\beta=0.01$, agree to within the stated uncertainties.  The lower panel of Figure \ref{mdfs} shows the fraction of PISN-dominated stars for $\beta=0.1$ (dark blue line) and $\beta=0.01$ (light blue line).  Evidently, the fraction also peaks significantly above the cut-off of the observed MDF at $[\mathrm{Ca}/\mathrm{H}]\sim -4$ and goes to zero approaching this cut-off. We have additionally run a simulation with iron as the reference element. The corresponding distribution of PISN-dominated stars (here defined such that $>90\%$ of the Fe is synthesized in PISNe) peaks at $[\mathrm{Fe}/\mathrm{H}]\simeq -2.5$, which concurs with the above result that these stars are to be found at relatively high metallicities. In this simulation, there is a weak tail of PISN-dominated stars towards low metallicities, originating from the least massive PISNe. However, for $\beta=0.01$, merely $\sim 1$ star in $1000$ show a clear PISN signature at $[\mathrm{Fe}/\mathrm{H}]=-4$, nearly a factor of $100$ less than around the peak at $[\mathrm{Fe}/\mathrm{H}]=-2.5$.

\subsection{Observational Constraints on Pop~III Supernovae}\label{obsconstr}
As it appears from Figure \ref{mdfs}, the fraction of PISN-dominated stars depends strongly on the parameter $\beta$, the fraction of Pop~III stars becoming PISNe, which is to be expected. This dependence may be used to put constraints on the number of PISN-dominated stars expected to be found in the metal-poor Galactic halo and, conversely, to put observational upper limits on $\beta$ itself. Figure \ref{apisn} shows the integrated fraction of PISN-dominated stars as a function of $\beta$. This fraction, termed $a_{\gamma\!\gamma}$, is defined as the number of PISN-dominated stars below $[\mathrm{Ca}/\mathrm{H}]=-2$ divided by the total number of stars below $[\mathrm{Ca}/\mathrm{H}]=-2$. The fractions displayed in Figure \ref{apisn} are based on the assumption that a distinct PISN signature is present over the entire PISN mass range.  In the reminder of the paper, we shall further explicitly assume that this signature is detectable in stars if, and only if, $>90\%$ of the atmospheric Ca abundance originates from the PISNe.  In Figure \ref{apisn}, we have also indicated the predicted fractions $a_{\gamma\!\gamma}$ corresponding to $99\%$ and $99.9\%$ PISN-enrichment (see also Table \ref{levelofpisnenrichment}).  These results will be briefly discussed later in this section.

The number of observed Galactic halo stars below $[\mathrm{Ca}/\mathrm{H}]=-2$ for which abundance analyses based on high-resolution spectroscopy exists is roughly $600$ (N. Christlieb, priv. comm.).  None of these $\sim 600$ stars appear to show a dominant PISN signature.  If so, we estimate that $\beta < 0.07$ (see Fig. \ref{apisn}), which corresponds to a mass fraction of very massive Pop~III stars of $\lesssim 40\%$.  Here, we disregard Draco $119$ (Fulbright et al. 2004\nocite{fulbright04}) and the group of field stars studied by Ivans et al. (2003\nocite{ivans03}) as PISN-dominated, second generation stars (see \S \ref{discussion}).  An upper limit of $\beta=0.07$ is consistent with the predicted range of $\beta$ derived by Greif \& Bromm (2006\nocite{gb06}) and the $\beta$ derived from the IMF by Padoan et al. (2007\nocite{padoan07}), which all fall in the range $0.005\lesssim \beta \lesssim 0.02$. Conversely, as indicated in Figure \ref{apisn}, this range corresponds to a range in $a_{\gamma\!\gamma}$ of $1.5\times 10^{-4}\lesssim a_{\gamma\!\gamma}\lesssim 5\times 10^{-4}$, which, at face value, means that $\sim3-10$ times as many metal-poor (i.e., $[\mathrm{Ca}/\mathrm{H}]\le -2$) Galactic halo stars need to be analyzed in order to find a single second generation star with $>90\%$ of its Ca originating from primordial PISNe. In \S \ref{discussion}, these results will be discussed further.  

\begin{table}[t]
\caption{Fraction of PISN-enriched stars as a function of level of PISN-enrichment}
  \label{levelofpisnenrichment}
  \begin{tabular}{l|c}
     \hline
     \hline
     \\*[-0.5em]
\hfill\footnotesize{Level of}\hfill{} & \hfill\footnotesize{Fraction of}\hfill{} \\
\hfill\footnotesize{PISN-enrichment\tablenotemark{a}}\hfill{} & \hfill\footnotesize{PISN-enriched stars\tablenotemark{b}}\hfill{} \\
     \\*[-0.5em]
     \hline
     \\*[-0.8em]
\footnotesize{$>\,\,\,1\%$} & \footnotesize{$6.97\times 10^{-2}$} \\*[0.1em]
\footnotesize{$>10\%$} & \footnotesize{$2.34\times 10^{-2}$} \\*[0.1em]
\footnotesize{$>50\%$} & \footnotesize{$3.60\times 10^{-3}$} \\*[0.1em]
\footnotesize{$>90\%$} & \footnotesize{$3.04\times 10^{-4}$} \\*[0.1em]
\footnotesize{$>99\%$} & \footnotesize{$7.35\times 10^{-5}$} \\*[0.1em]
\footnotesize{$>99.9\%$} & \footnotesize{$6.75\times 10^{-5}$} \\
     \\*[-0.8em]
     \hline
     \\*[-0.8em]
  \end{tabular}

\tablenotetext{a}{As measured by the amount of Ca originating from PISNe.}
\tablenotetext{b}{All fractions are calculated for $\beta=0.01$.}
\end{table}

Also shown in Figure \ref{apisn} are the predicted fractions of stars that are pre-enriched to at least $99\%$ (medium sized, blue dots) and $99.9\%$ (small, dark blue dots) by PISNe. These will be referred to as purely PISN-enriched stars. Although the number of such stars is smaller than the number of stars in which up to $10\%$ of the metals (Ca) are allowed to originate from core collapse SNe, it is not radically smaller. In the range $0.005\lesssim \beta \lesssim 0.02$, e.g., the predicted fraction of purely PISN-enriched stars is only smaller by factor of $\sim 4$ (see also Table \ref{levelofpisnenrichment}). Moreover, assuming that none of the $\sim 600$ stars below $[\mathrm{Ca}/\mathrm{H}]=-2$, for which high-resolution spectroscopic data are available, are purely PISN-enriched stars, the range of possible values for $\beta$ is constrained to be $<0.2$. As indicated above, a slightly stronger constraint may be placed on $\beta$ if the reasonable assumption is being made that none of these stars has a PISN-enrichment which exceeds $90\%$. However, due to the increasing risk of making false identifications, $\beta$ may not be constrained much further by allowing for lower levels of PISN-enrichment than $90\%$.  Table \ref{levelofpisnenrichment} shows the predicted fraction of stars formed from gas pre-enriched by PISNe to various levels, for $\beta=0.01$. Clearly, the fraction of PISN-enriched stars increases with decreasing level of PISN-enrichment.  However, even though the expected number of stars with, e.g., a PISN-enrichment of $>50\%$ is $\sim 10$ times more than the corresponding number with $>90\%$ PISN-enrichment, we would in practice not gain much as it would most likely be very difficult to correctly identify many of the less PISN-dominated stars.  Note also that for $\beta=0.01$, less than $10\%$ of all stars below $[\mathrm{Ca}/\mathrm{H}]=-2$ are formed from gas enriched to any level by PISNe.

\subsection{Parameter Dependence and Sensitivity}\label{uncertainties}
We have performed a number of simulations where we varied the values of several model parameters, in order to illustrate the sensitivity of our results to the uncertainties in these parameters. The SFR is a crucial ingredient in the model and we have varied both the Pop~III and Pop~II SN rates by changing $u_{\mathrm{I\!I\!I,0}}$ and $u_{\mathrm{I\!I,0}}$, respectively. The turbulent diffusion coefficient $D_{\mathrm{t}}$ in the early ISM is difficult to estimate. According to Bateman \& Larson (1993\nocite{bl93}), the diffusive mixing of matter in the cold, neutral medium in the Galactic disk (e.g., through molecular cloud collisions) occurs on a similar time scale as the one adopted here, while the mixing should be substantially faster in hot, ionized media. Alternatively, Pan \& Scalo (2007\nocite{pan07}) developed a "slow" mixing model with inefficient (i.e., low $D_{\mathrm{t}}$) diffusion to explain the apparent existence of primordial gas at relatively low ($z\sim 3$) redshift. We have varied $D_{\mathrm{t}}$ in our model to map out the effect. 

\begin{table}[t]
\caption{\footnotesize{The dependence of $a_{\gamma\!\gamma}$ on model parameters.  The three parameters which were varied are listed in the top row; from left to right, these are the Pop III SN rate, the Pop II SN rate, and the turbulent diffusion coefficient.  The leftmost column shows the factor by which these parameters were changed from their fiducial values, and the resulting values for $a_{\gamma\!\gamma}$ are shown in the three rightmost columns below the specific parameter which was varied.}}
  \label{parameters}
  \begin{tabular}{l|ccc}
     \hline
     \hline
     \\*[-0.5em]
\hfill\footnotesize{Parameter\tablenotemark{a}}\hfill{} & \hfill\footnotesize{$u_{\mathrm{I\!I\!I,0}}$}\hfill{} & \hfill\footnotesize{$u_{\mathrm{I\!I,0}}$}\hfill{} & \hfill\footnotesize{$D_{\mathrm{t}}$}\hfill{} \\
     \\*[-0.5em]
     \hline
     \\*[-0.8em]
\footnotesize{$\times 10$} & \footnotesize{$1.07\times 10^{-4}$} & \footnotesize{$7.35\times 10^{-4}$} & \footnotesize{$6.89\times 10^{-5}$} \\*[0.1em]
\footnotesize{$\times \,\,\,3$} & \footnotesize{$1.57\times 10^{-4}$} & \footnotesize{$1.99\times 10^{-4}$} & \footnotesize{$7.98\times 10^{-5}$} \\*[0.1em]
\footnotesize{$\times \,\,\,1$\tablenotemark{b,c}} & \footnotesize{$3.04\times 10^{-4}$} & \footnotesize{$3.04\times 10^{-4}$} & \footnotesize{$3.04\times 10^{-4}$} \\*[0.1em]
\footnotesize{$\times \,\,\,1/3$} & \footnotesize{$4.78\times 10^{-4}$} & \footnotesize{$4.59\times 10^{-4}$} & \footnotesize{$1.02\times 10^{-3}$} \\*[0.1em]
\footnotesize{$\times \,\,\,1/10$} & \footnotesize{$4.01\times 10^{-4}$} & \footnotesize{$3.41\times 10^{-4}$} & \footnotesize{$1.92\times 10^{-3}$} \\
     \\*[-0.8em]
     \hline
     \\*[-0.8em]
  \end{tabular}

\tablenotetext{a}{The fiducial values of the model parameters are discussed in \\\S \ref{ece}.  All fractions are calculated for $\beta=0.01$.}
\tablenotetext{b}{$a_{\gamma\!\gamma}=4.78\times 10^{-4}$, adopting the set of PISN yields calculated \\by Umeda \& Nomoto (2002).}
\tablenotetext{c}{$a_{\gamma\!\gamma}=6.03\times 10^{-4}$, assuming that $\mathrm{SFR}\propto \rho^2$}
\end{table}

The results of the simulations are summarized in Table \ref{parameters}, which shows the dependence of $a_{\gamma\!\gamma}$ on $u_{\mathrm{I\!I\!I,0}}$, $u_{\mathrm{I\!I,0}}$, and $D_{\mathrm{t}}$, respectively. In all simulations, $\beta$ is fixed to $0.01$.  The impact on $a_{\gamma\!\gamma}$ when altering the Pop~III (col. $1$) and Pop~II (col. $2$) SN rates is non-trivial, with, respectively, a local maximum and a local minimum in the value of $a_{\gamma\!\gamma}$ appearing within the examined ranges of the values for these rates. The dependence on $D_{\mathrm{t}}$ (col. $3$) is less complex. We shall not discuss these dependences in detail but notice that $a_{\gamma\!\gamma}$ is roughly proportional to \mbox{$\int u_{\mathrm{I\!I\!I}}\,u_{\mathrm{I\!I}}\mathrm{d}t\,/\!\int u_{\mathrm{I\!I}}^2\mathrm{d}t$} and that the Pop~III and Pop~II rates are coupled through $Q_{\mathrm{p}}$ (see eqs. (\ref{pop3eq})--(\ref{muiii})). In particular, the counter-intuitive decrease in $a_{\gamma\!\gamma}$ with an increasing Pop~III SN rate above $u_{\mathrm{I\!I\!I,0}}=1.33\times 10^{-2}$ kpc$^{-3}$ Myr$^{-1}$ ($=1/3$ of the adopted $u_{\mathrm{I\!I\!I,0}}$) may be understood from the fact that an initially high Pop~III rate more efficiently suppresses subsequent formation of Pop~III stars, while it simultaneously promotes the formation of Pop~II stars, as a consequence of the change in the fraction of primordial gas. Note that a similar effect occurs when $\beta$ is increased, indicated in Figure \ref{pop3}.

The overall change in $a_{\gamma\!\gamma}$ over the entire two decade range in $u_{\mathrm{I\!I\!I,0}}$, as well as in $u_{\mathrm{I\!I,0}}$, is merely a factor of $\sim 4$, which is quite moderate. The sensitivity on $D_{\mathrm{t}}$ is higher, a factor of $\sim 30$ over the examined range. We also ran one simulation where the SFR was set proportional to the square of the density instead of the normal linear dependence, and one where we changed the set of PISN yields to those by Umeda \& Nomoto (2002\nocite{un02}). In each of these two runs, $a_{\gamma\!\gamma}$ changed by less than a factor of $2$ (see footnote $\mathrm{b}$ and $\mathrm{c}$ in Table \ref{parameters}).

The data presented in Table \ref{parameters} are a measure of the uncertainty in our calculations. Evidently, the result is insensitive to the SFR and PISN yields, while it is relatively sensitive to the amount of mixing in the ISM.  If we, e.g., adopt a $10$ times smaller turbulent diffusion coefficient, the fraction of PISN-dominated stars is increased by a factor of $6$.  Still, even for such a low $D_{\mathrm{t}}$, the range of $\beta$ predicted by Greif \& Bromm (2006\nocite{gb06}) would be consistent with the observational upper limit.  We would like to emphasize that in all the runs that we have carried out, the vast majority of the PISN-dominated stars have significantly higher Ca abundances than the most Ca-poor stars in the simulations, irrespective of the predicted value of $a_{\gamma\!\gamma}$.  Moreover, for all the parameters that we have varied, $\beta$ remains the parameter to which the results are most sensitive (cf. Fig. \ref{apisn}). The above findings indicate two points: first, due to the sensitivity on $\beta$, we have the means to measure the frequency of very massive stars in the early universe from observations of large, homogeneous samples of metal-poor stars in the Milky Way halo.  Second, due to the insensitivity on model parameters apart from $\beta$, our results and the conclusions based on these results should be relatively robust.

\section{Discussion and Summary}\label{discussion}
We have shown that Milky Way halo stars born in atomic cooling halos of \mbox{$M\sim10^8~\mathcal{M_{\odot}}$} and displaying a chemical abundance pattern characterized by that of Pop~III PISNe are to be found at significantly higher metallicities than where the most metal-poor stars have been found. Consequently, although stars in the metal-poor tail of the MDF, like the extremely iron-deficient stars HE $0107$-$5240$ (Christlieb et al. 2002\nocite{cetal02}) and HE $1327$-$2326$ (Frebel et al. 2005\nocite{frebel05}), are very important for the understanding of the star formation process (e.g., Bromm \& Loeb 2003\nocite{bromm03}; Umeda \& Nomoto 2003\nocite{un03}; Tumlinson 2007\nocite{tumlinson07}) and feed-back mechanisms (e.g., Karlsson 2006\nocite{karlsson06}) in the early universe, they seem to tell us little about primordial PISNe. The majority of Pop~II stars with a dominant contribution from PISNe are predicted to have $\mathrm{Ca}$ abundances in excess of $[\mathrm{Ca}/\mathrm{H}]=-2.6$, the upper limit for which the Hamburg/ESO survey of metal-poor stars is considered to be complete. This implies that a significant fraction of the PISN-dominated stars may have escaped detection, not because they are too $\mathrm{Ca}$-poor but because they are too $\mathrm{Ca}$-rich. The number fraction of very massive Pop~III stars exploding as PISNe is estimated to be $\beta < 0.07$ ($\lesssim 40\%$ by mass), assuming that currently no Galactic halo star below $[\mathrm{Ca}/\mathrm{H}]=-2$ with available detailed abundance analysis shows a chemical signature characteristic of PISNe.  This result is consistent with the theoretical prediction of $0.005\lesssim \beta \lesssim 0.02$, which corresponds to a handful of PISN-dominated, second generation stars in a sample of $10~000$. The number of purely PISN-enriched stars is predicted to be a factor of $\sim 4$ smaller, assuming this range of $\beta$.

In an interesting paper, Salvadori et al. (2007\nocite{salvadori07}) also discuss the existence of second generation stars and their location in the MDF. Their stochastic model of chemical enrichment differs from ours in a number of aspects. Firstly, they model the chemical evolution of the Galaxy as a whole based on the merger-tree approach (cf. Tumlinson 2006\nocite{tumlinson06}), while we focus on the evolution of atomic cooling halos using the mixing volume picture. They also assume that all Pop~III stars are very massive, while we allow for less massive Pop~III star formation as well, controlled by the parameter $\beta$. Hence, in their models with $Z_{\mathrm{crit}}>0$ (for $Z_{\mathrm{crit}}=0$, no PISNe are formed), all second generation stars are formed from gas only enriched by PISNe.  Finally, we note that while our chemical enrichment model accommodates the important contribution of metals ejected by the supernovae from Pop~III stars formed in minihalos, with masses less than 10$^8$ $M_{\odot}$ (see Bromm et al. 2003\nocite{byh03}; Greif et al. 2007\nocite{greif07}), this contribution is neglected in Salvadori et al. (2007\nocite{salvadori07}) based on the assumption that strong photodissociating radiation from the first stars quenches their formation. It appears likely, however, that this effect may not have substantially lowered the Pop~III star formation rate (e.g. Ricotti et al. 2001\nocite{ricotti01}; Johnson et al. 2007\nocite{jgb07}; O'Shea \& Norman 2007\nocite{oshea07}; Wise \& Abel 2007\nocite{wise07}; see also Yoshida et al. 2003\nocite{yoshida03}).  Despite these differences, both models generate broadly similar MDFs.  However, at variance with the results presented here, Salvadori et al. predict that the highest fraction of PISN-dominated stars is to be found in the lowest metallicity regime and only a very small fraction at $[\mathrm{Fe}/\mathrm{H}]> -1$, provided that the host halo is able to form both first and second generation stars.  This finding may be a consequence of these authors' assumption of the instantaneous mixing of metals in the galactic medium, but should be analyzed in more detail to be fully understood.

The most characteristic and reliable chemical signature of PISNe is probably the absence of neutron-capture elements. Although this may not be a unique fingerprint (massive core collapse SN may be unable to synthesis n-capture elements as well), we have shown that stars formed from material pre-enriched by PISNe emerge in a metallicity regime where both the r-process and the s-process signatures are thoroughly established (see Burris et al. 2000\nocite{burris00}). The presence of n-capture elements in normal Pop~II stars will thus serve to rule out PISN pre-enrichment.  Metal-poor stars with enhanced abundances of r-process elements are certainly interesting in their own right as they, e.g., can be used as cosmochronometers and allow us to study the origin and nature of the r-process.  However, as important as it is to search for r-process enhanced stars (see e.g., Christlieb et al. 2004\nocite{christlieb04}), alternatively, the search for stars unusually depleted in r-process elements may help us locate candidates of the ``missing'' second stellar generation.  With high-resolution follow-up spectroscopy, further signatures indicative of PISNe such as a pronounced odd-even effect and, e.g., an atypical $\mathrm{Fe}/\mathrm{Ca}$ ratio (see Fig. \ref{fecah}), should then be measured to ultimately verify or refute candidates as PISN-dominated, second generation stars. Although the actual number of PISN-dominated stars is predicted to be small, perhaps only one or two in $10~000$ below $[\mathrm{Ca}/\mathrm{H}]=-2$, the ongoing and future large-scale surveys, such as the SDSS\footnote{The Sloan Digital Sky Survey}, SEGUE\footnote{The Sloan Extension for Galactic Understanding and Evolution. Operating since 2005.}, and LAMOST\footnote{The Large sky Area Multi-Object fiber Spectroscopic Telescope. First spectrum delivered May 28, 2007.}, aim to find at least an order of magnitude more stars in this metallicity regime than is presently known. A few thousand of these stars should be bright enough to be successfully observed with high-resolution by $8-10$ m class telescopes (Beers \& Christlieb 2005\nocite{bc05}). With the advent of $30-100$~m class telescopes, high-resolution spectroscopic follow-up will become feasible for the majority of stars discovered below $[\mathrm{Ca}/\mathrm{H}]=-2$.

There are a few examples of metal-poor stars in the Galactic halo which do show anomalous abundance patterns. Ivans et al. (2003\nocite{ivans03}) studied a group of three dwarf/subgiant stars with $[\mathrm{Ca}/\mathrm{H}]\simeq-2.2$, BD $80^{\circ}~245$, G $4$-$36$, and CS $22966$-$043$, which appears to be heavily depleted in n-capture elements (see also, e.g., HD $122563$, recently studied by Mashonkina et al. 2008\nocite{mashonkina08}).  These stars also display low abundances of $\mathrm{Na}$, $\mathrm{Al}$, and $\alpha$-elements, as indicated by the relatively high $[\mathrm{Fe}/\mathrm{Ca}]$ ratios in Figure \ref{fecah}. Interestingly, BD $80^{\circ}~245$ and G $4$-$36$ both show a particularly low $[\mathrm{Al}/\mathrm{Ca}]$ ratio and BD $80^{\circ}~245$ and CS $22966$-$043$ have $[\mathrm{Mg}/\mathrm{Ca}]<0$. However, the $[\mathrm{Na}/\mathrm{Ca}]$ ratio in all three stars is found to be typical for Galactic halo stars and although the heavy iron-peak elements are low in BD $80^{\circ}~245$, they are strongly enhanced in G $4$-$36$, and CS $22966$-$043$, an observation which is not understood. Ivans et al. (2003\nocite{ivans03}), explain the peculiar abundance patterns as a result of an early contribution by thermonuclear SNe (Type Ia) in the subsystems in which these stars were born. They do not consider PISNe as an alternative, presumably because very massive stars traditionally are believed to contribute to the chemical enrichment only at the very lowest metallicities. As intriguing as it is to envision these objects, in particular BD $80^{\circ}~245$, as PISN-enriched, second generation stars, the match is ambiguous. Further investigation should, however, be conducted on these peculiar stars. Another star shown in Figure \ref{fecah} is Draco $119$. This $[\mathrm{Fe}/\mathrm{H}]=-2.9$ giant is a member of the Draco dSph and is the only star currently known to be completely devoid of elements above $\mathrm{Ni}$, to within detection limits (Fulbright et al. 2004\nocite{fulbright04}). The upper limit on the $\mathrm{Ba}$ abundance, e.g., is estimated to $[\mathrm{Ba}/\mathrm{Fe}]<-2.6$, more than a full dex below the next most $\mathrm{Ba}$-poor stars at $[\mathrm{Fe}/\mathrm{H}]\sim -3$. Based on the light element abundances, Fulbright et al. (2004\nocite{fulbright04}) conclude that Draco $119$ was born out of gas primarily enriched by massive core collapse SNe, and not by PISNe, which is consistent with our finding that PISN-dominated stars generally should appear above this metallicity regime.

We have assumed that the metal-poor Galactic halo mainly consists of stars formed during the assembly of atomic cooling halos. We note, however, that a significant fraction of these halos may have ended up in the bulge of the Milky Way. With the new high-resolution infrared spectrographs on the $8-10$ m class telescopes, detailed spectral analysis of dust-obscured Galactic bulge stars has recently become feasible (e.g., Ryde et al. 2007\nocite{ryde07}). The search for relics of the first stars should therefore be pursued also in this, previously inaccessible part of our Galaxy.

\acknowledgements
V. B. acknowledges support from NSF grant AST-0708795. The cosmological simulation presented here was carried out at the Texas Advanced Computing Center (TACC).  We would like to thank Anna Frebel, Norbert Christlieb, Kjell Eriksson, Stefania Salvadori, and Raffaella Schneider for helpful discussions.  Finally, we would like to acknowledge the anonymous referee whose comments have greatly improved the presentation of this work.

\bibliographystyle{bibtex/apj}
\bibliographystyle{bibtex}
\bibliography{ref}

\end{document}